\documentstyle[preprint,pra,aps]{revtex}
\begin{document}
\draft
\title{\bf
Effects of T- and P-odd weak nucleon
interaction in nuclei:\\
renormalizations due to residual strong interaction,
matrix elements between compound states and their correlations
with P-violating matrix elements
\footnote{{\bf Submitted to PHYSICAL REVIEW C}}}
\author{V.~V.~Flambaum and O.~K.~Vorov}
\address{School of Physics, University of New South Wales,
Sydney 2052, New South Wales, Australia}
\date{28 November 1994}
\maketitle
\begin{abstract}
Manifestations of P-,T-odd weak interaction between nucleons in nucleus
are considered. Renormalization of this interaction due to
residual strong interaction is studied. Mean squared matrix
elements of P-,T-odd weak interaction between compound
states are calculated. Correlators between P-,T-odd and P-odd, T-even
weak interaction matrix elements between compound states are
considered and estimates for these quantities are obtained.
\end{abstract}
\vspace{10mm}
\narrowtext

\section{Introduction}
\label{sec:level1}

Spatial parity nonconserving weak interaction of nucleons is now
subject for extensive experimental and theoretical investigations.
In these studyings, quantitative comparison of experimental results
and theoretical predictions is possible.
Developments in experimental techniques
and interpretations of the results thus obtained allow one
to rise questions going far from the scope of theory of weak
interaction \cite{FRANKLE}-\cite{MITCHEL}.

Much less is known, both in experimental and theoretical  aspects,
about the component of nucleon weak interaction that violates
both spatial parity (P) and time reversal invariance (T)
(T-,P-odd weak interaction).
The problem of possible T-violation
has been of interest for a long time \cite{MOLDAUER}-\cite{BLINSTOYLE}.
In the context of nuclear physics,
the T-,P-odd interaction, if exists,
induces T-,P-odd nuclear moments
\cite{HAXTON}-\cite{PHI}
(electric dipole, magnetic quadrupole moments, ``Schiff'' moment {\it etc}).
Experimental data exist only for upper
limits of these quantities.
At the same time, theoretical values of constants of this
interaction are not unambiguously known, varying by several
orders of magnitude from one model to another
(see e.g., \cite{FKSNP}, \cite{PHI}).
In most cases, the scale of T-,P-odd interaction is predicted
to be very small.

In this situation, possible sources of enhancement of the effects
caused by this interaction, which allow experimental investigation,
are crucial for the further studies of T-,P-odd interaction.
Apparently, the compound nuclear resonances providing a large statistical
enhancement of small perturbations, are very convenient in this
case.

We are considering here the weak T-,P-odd nucleon interaction in nuclei
beginning at the single-particle level.
The effects of the residual strong interaction on the T-,P-odd potential
was considered in Ref. \cite{PION}.
In present work, we focus attention on renormalization
of the two-body T-,P-odd interaction
due to the residual strong interaction,
which is important for description of the T-,P-odd effects in nuclear
states at excitation energies near or lower than that of neutron separation
threshold, $B_{n}$.
We have calculated
mean squared  T-,P-violating matrix elements between compound
states and have considered possible correlations of these matrix elements
with the
matrix elements of P-odd, T-even weak interaction \cite{we}.

The structure of the paper is following.
In section II we consider T-,P-odd potential, acting on a nucleon
that arises in mean field approximation for the initial two-body
weak interaction. We calculate the single-particle matrix elements
of this potential and discuss their properties in comparison with the
single-particle matrix elements of P-odd, T-even weak interaction.
In section III we consider renormalization of the T-,P-odd weak interaction
by residual strong nucleon interaction.
In section IV the equations of this renormalization are solved for
the Landau-Migdal parametrization of the residual strong interaction
and explicit analytical results for the effective two-body T-,P-odd weak
interaction between the nucleons in heavy nuclei are derived.
Numerical results
are obtained for the matrix elements expressed through the constants
of the initial weak interaction. It is shown that, in contrary to the case
of P-odd, T-even weak interaction, renormalization due to strong
interaction does not result in enhancement of matrix elements, though
this renormalization is important for quantitative results.

In section V we calculate mean squared matrix elements of the
P-,T-odd interaction between nuclear compound states of opposite parity
within statistical model. Section VI is devoted to discussion of
correlations of P-,T-odd  and P-odd, T-even matrix elements
between compound states. Calculation of the correlator in the
statistical model yields the value about 10 per cent.

The results are summarized
in section VII.

\section{T-,P-odd weak nucleon interaction. T-,P-odd potential}
\label{sec:level2}

The nuclear Hamiltonian $H$ with account for the T-,P-odd
weak interaction can be written in the form
\begin{equation}
H=H_{0} \quad + \quad V_{S} \quad +
\quad {\cal W}^{{\small P,T}} + \quad F,
\end{equation}
Here, the first term $H_{0}={\bf p}^{2}/2m+U_{S}(r,\vec{\sigma})$ is the single
particle
Hamiltonian of the nucleons moving in the strong mean field
$U_{S}(r,\vec{\sigma})$ including the spin-orbit interaction,
$V_{S}$ stands for the residual
two-body strong interaction
(it will be considered in section IV),
and ${\cal W}^{{\small P,T}}$ describes
P-,T-odd weak interaction between nucleons,
$F$ denotes other possible interactions, e.g., coupling to
electromagnetic field.
The two-body weak P-,T-odd interaction
${\cal W}^{{\small P,T}}$
can be written
as follows
(see e.g. \cite{TP1},\cite{FKSNP}):

\begin{eqnarray}
\hat {\cal W}^{{\small P,T}}(1,2)  =
\frac{G}{\sqrt{2}} \frac{1}{2m}
\Bigl(
(\eta_{12} \vec{\sigma}_{1}- \eta_{21} \vec{\sigma}_{2}) \cdot
\vec{\nabla}_{1} \delta(\vec{r}_{1}-\vec{r}_{2})+
                                     \nonumber\\
\eta'_{12} [\vec{\sigma}_{1} \times \vec{\sigma}_{2}] \cdot
\lbrace \vec{p}_{1}-\vec{p}_{2},
\delta(\vec{r}_{1}-\vec{r}_{2}) \rbrace
\Bigl)
\end{eqnarray}
where $G=10^{-5}m^{-2}$ is the Fermi constant, $m$ is the nucleon mass,
${\bf p}$ and $\vec{\sigma}$ are the nucleon momentum and doubled spin
respectively. Hereafter,
$\vec{a} \times \vec{b}$
means exterior vector product, and $\{a,b \}$ denotes anticommutator.
The dimensionless constants $\eta_{1,2},\eta'_{1,2}$ which determine
the scale of the T-,P-odd effects, are predicted
to be very small, e.g., within Kobayashi-Maskawa model
(see e.g.,\cite{TP1},\cite{FKSNP},\cite{FLMOD}).

The analysis of the P-,T-odd effects in nuclei is similar to
that in the case of the P-odd weak interaction
(see e.g. \cite{we},\cite{IPNCI}).
It is convenient to introduce P-,T-odd ``weak potential''
$w^{{\small P,T}}$ acting on a valence nucleon
$1$,
that arises from
summation ${\cal W}^{{\small P,T}}(1,2)$
over the states  of the nucleon $2$
(see , e.g. Ref.\cite{FKSNP}):
\begin{equation}
w^{{\small T,P}}=\frac{G}{2 \sqrt{2} m} \eta (\vec{\sigma} \vec{\nabla})
\rho(r)
\end{equation}
where $\rho$ is the nucleon density,
the dimensionless constants $\eta_{p},\eta_{n}$
characterize the strength of T- and P-odd potential for proton
(neutron); they are connected to the parameters of the initial two-body
interaction ${\cal W}^{{\small P,T}}(1,2)$ by the relations
\begin{equation}
\eta_{p}=\frac{Z}{A}\eta_{pp} + \frac{N}{A}\eta_{pn},
\quad
\eta_{n}=\frac{N}{A}\eta_{nn} + \frac{Z}{A}\eta_{np},
\end{equation}
where $Z,N$, and $A$ are the nuclear charge, a neutron number and
its mass number respectively.
The limits on these constants ($\eta_{p},\eta_{n}$)
were obtained from the atomic \cite{ATOM} and molecular \cite{MOLECULE}
electric dipole
moment measurements.
Being a single-particle operator, the T-,P-odd weak potential
$w^{{\small T,P}}$ obeys the same selection
rules as the P-odd, T-even weak potential $w^{{\small P}}$:
\begin{equation}
\Delta l = \pm 1, \quad \Delta j = 0.
\end{equation}
The values of the matrix elements of the P-,T-odd weak potential (3) between
single-particle nuclear states calculated for $^{209}Pb$ are presented
in Tables I,II.
The numerical calculations
have been performed with the use the
of single-particle basis of states obtained by numerical solution
of the eigenvalue
problem in the Woods-Saxon potential with spin-orbital interaction  in the
form
\begin{equation}
U_{S}(r,\vec{\sigma})=
-U_{0}f(r)+U_{ls}({\bf \sigma l})(\hbar/(m_{\pi}c))^{2}\frac{1}{r}
\frac{df}{dr}+ U_{c}
\end{equation}
with $f(r)=(1+exp((r-R)/a))^{-1}$.
Here,
${\bf l}$ is the orbital angular momentum, $U_{c}$ means
Coulomb correction for protons, $U_{c}=3Ze^{2}/(2R)(1-r^{2}/(3R^2)), r\leq R$
and $U_{c}=Ze^{2}/r, r>R$, for $R$, $a$, and $r$ being the nuclear radius,
diffusity parameter and radial variable correspondingly. The parameter values
were used in accordance with Bohr-Mottelson formulas (see Ref.\cite{BM})
for the
case of $^{233}Th$: they are close to those established
for heavy nuclei like
lead (Ref.\cite{Brown}) to reproduce single-particle properties.

As it is seen from Tables I,II, the single-particle T-,P-odd matrix elements
(column 5) are numerically suppressed (by about order of magnitude) as
compared to the matrix elements of P-odd, T-even potential
$\hat w^{P}(1) = \langle  \hat W^{P}(1,2) \rangle =
\frac{Gg}{2\sqrt{2}m}\lbrace({\bf \sigma}{\bf p})\rho +
\rho ({\bf \sigma}{\bf p})\rbrace$ that arises from the
corresponding two-body interaction
\cite{HAXTON}
\begin{equation}
\hat W (1,2)  = \frac{G}{\sqrt{2}} \frac{1}{2m}
\Bigl( (g_{12}{\bf \sigma}_{1}- g_{21}{\bf \sigma}_{2}) \cdot \lbrace
({\bf p}_{1}-{\bf p}_{2}), \delta({\bf r}_{1}-{\bf r}_{2}) \rbrace
+
g'_{12} [{\bf \sigma}_{1} \times {\bf \sigma}_{2}] \nabla_{1}
\delta({\bf r}_{1}-{\bf r}_{2}) \Bigl)
\end{equation}
in the same way as the P-,T-odd potential.
The differnce between these two cases is due to surficial
character of the potential (3) that is proportional
to the nuclear density derivative and peaked at nuclear surface.
As average, the mixing of the single-particle states of opposite
parity due to potential (3) that give rise to nuclear T-,P-odd nuclear
moments (see \cite{TP1},\cite{FKSNP})
\begin{equation}
f_{ab}=w^{TP}_{ab}/(\varepsilon_{a}-\varepsilon_{b}) \sim
10^{-8}\eta_{12}.
\end{equation}

Due to the selection rules (5), the following considerations,
analogous to those in the case of P-odd, T-even weak interaction,
take place here. It is well known
\cite{Zar}, \cite{KADMENSKY}
that doublets of single-particle states
with the same total angular momentum, but of opposite parity usually
do not appear in the same spherical nuclear shell. Thence, the energy
separation between levels in such doublets is about $5...8 MeV$,
the average energy distance
between different shells.
Thus, the coherent single-particle P-,T-odd contribution (3) does not work
effectively in mixing of  any excited nuclear states
(including the compound states)
with the energies below
$S_{n}...4 \div 6 MeV$, the neutron separation energy,
because the many-particle wave functions in this energy region
are dominated by
nucleon excitations within the valence shells \cite{BM}.
Therefore, the main P-odd effects in this energy region are to be determined
by the purely two-particle ``residue'', $:\hat W(1,2):$ of
the weak interaction
$\hat W(1,2)$, given by the
difference
\begin{equation}
:{\cal W}^{{\small P,T}}(1,2): \quad \equiv
{\cal W}^{{\small P,T}}(1,2) -  \langle {\cal W}^{{\small P,T}}(1,2) \rangle
= {\cal W}^{{\small P,T}}(1,2) -  w^{{\small P,T}}(1),
\end{equation}
which does not contain coherent summation in contrast to (3).

We consider first the case with the strong interaction
$V_{S}$ being ``switched off''.
Technically,
it is convenient to include the corrections caused by
the T,P-odd potential (5) into
the single-particle wave functions using unitary transformation.
As is known from Refs.\cite{TP1},\cite{FKSNP},
in the simple model with the strong potential $U(r)$ being proportional
to the nuclear density $\rho$ ($\rho(r)=\rho(0) U(r)/U(0)$),
it is easy to find the result of the action of the perturbation
$\hat w(1)$
\begin{eqnarray}
\tilde \psi =
exp(-{\hat \alpha}) \psi^{0}
\simeq (1-\theta \vec{\sigma} \vec{\nabla} )\psi^{0},
\qquad {\hat \alpha}=\theta \vec{\sigma} \vec{\nabla}
                              \nonumber\\
\theta = \eta \frac{G}{2 \sqrt{2} m} \frac{\rho(0)}{U(0)}=
-2 \cdot 10^{-8}
\eta \cdot fm,
\end{eqnarray}
where $\psi^{0}$ is the unperturbed wave function, and $\tau_{z}=-1(+1)$ is
isospin projection for proton(neutron). To get this solution, one
should also neglect spin-orbit interactions.
Accordingly, the matrix elements of any operator $O$, including the
Hamiltonian, can be calculated by using the unperturbed wave functions
$\psi^{0}$ and the transformed operator $\tilde O$:
\begin{displaymath}
\langle \tilde \psi_{a}|O|\tilde \psi_{b} \rangle=
\langle  \psi_{a}^{0}|\tilde O| \psi_{b}^{0} \rangle=
\langle  \psi_{a}^{0}|e^{\hat \alpha} O e^{-\hat \alpha}| \psi_{b}^{0} \rangle
\simeq \langle  \psi_{a}^{0}|O + [{\hat \alpha},O]| \psi_{b}^{0} \rangle,
\end{displaymath}
where $e^{\hat \alpha} \equiv e^{i \theta ( \vec{\sigma} \vec{\nabla})}$
is the operator
of the corresponding unitary transformation with the single-particle
anti-Hermitian $\hat \alpha$. This transformation
compensates the single-particle P-,T-odd potential in the Hamiltonian
$e^{\hat \alpha} H e^{-\hat \alpha}$.
The effect of this potential is now included into the renormalized operators
${\tilde O}$ rather than the wave functions ${\tilde \psi}$.

\section{Renormalization of the P-,T-odd
effects due to residual strong interaction}
\label{sec:level3}

To take the strong interaction $V_{S}$ into account, let us
seek now for an operator $e^{{\hat {\cal A}}}$
which should play the same role as $e^{{\hat \alpha}}$ above,
but will incorporate the renormalization effects due to
the residual strong interaction $V_S$.
Eventually, as we will see below the operator ${\hat {\cal A}}$
differs from ${\hat \alpha}$ mainly
due to the renormalization
of the weak interaction constant by the residual strong interaction $V_{S}$.
The transformed Hamiltonian looks like:
\begin{eqnarray}
\tilde H \quad = \quad e^{{\cal A}}H e^{-{\cal A}} \quad =
H_{0} \quad + \quad V_{S} \quad + \quad \hat F \quad +
\qquad \qquad
\nonumber\\
+ \quad w^{P,T} \quad + \quad :\hat {\cal W}^{P,T}:  \quad
+\quad [{\hat {\cal A}},H_{0}] \quad +
\quad  [{\hat {\cal A}},V_{S}] \quad + \quad [{\hat {\cal A}},F] \quad
\end{eqnarray}
where we have used the decomposition (4) and neglected all terms above
the first order in the weak interaction.
To obtain the effective two-particle P,T-odd interaction acting
in the valence shells we should find  the
operator ${\hat {\cal A}}$ in such a way that the single-particle
P,T-odd contribution
in $e^{\hat {\cal A}}He^{-{\hat {\cal A}}}$ will be compensated.
The last term in (6) is
a two-body operator. We employ the same decomposition, as in (4):
$[{\hat {\cal A}},V_{S}] \equiv \langle [{\hat A},V_{S}] \rangle +
:[{\hat {\cal A}},V_{S}]:$, where the
first single-particle term is the average over the paired nucleons,
and the second one, $:[{\hat {\cal A}},V_{S}]:$,
which yields zero under such averaging,
is the effective induced two-particle interaction which we are seeking
for:
\begin{equation}
{\cal W}^{TP}_{ITPNCI}=
\quad :[{\hat A},V_{S}]: \quad , \qquad \langle {\cal W}^{TP}_{ITPNCI} \rangle
\equiv 0.
\end{equation}
Now we choose the operator ${\hat {\cal A}}$ in such a way
that the ``compensation equation''
\begin{equation}
\hat w^{{\small P,T}}  \quad + \quad [{\hat {\cal A}},H_{0}] \quad +
\quad \langle [{\hat {\cal A}},V_{S}] \rangle =0,
\end{equation}
is fulfilled. After that, the transformed Hamiltonian (11)
takes the form
\begin{eqnarray}
\tilde H = H_{0} \quad + \quad V_{S} \quad + \quad F \quad +
\quad :\hat {\cal W}^{ P,T}:
\nonumber\\
\quad +
\quad {\cal W}^{P,T}_{ITPNCI} \quad + \quad [{\hat {\cal A}},F]
\end{eqnarray}
where P-,T-odd single-particle terms are canceled.
The sources of symmetry violations presented in Eq.(16) can be
classified as follows:

(i) the term $[{\hat {\cal A}},F]$
that gives a direct contribution
of the symmetry violating potential
$w^{{\small P,T}}_{1}$
to the matrix elements of an external field $F$
( $\langle \psi | F + [{\hat {\cal A}},F] | \psi ' \rangle =
\langle {\tilde \psi} | F | {\tilde \psi '} \rangle$);

(ii) The two-body residual weak interaction
$:{\cal W}^{{\small P,T}}:$ ;

(iii) ${\cal W}^{{\small P,T}}_{ITPNCI} $,
which play the same role as
$:{\cal W}^{{\small P,T}}:$. We note that the induced P-,T-odd
interaction
${\cal W}^{{\small P,T}}_{ITPNCI}$ is
not enhanced in comparison with the
two-particle
residual P-,T-odd interaction
$:{\cal W}^{{\small P,T}}:$, contrary to the case of P-odd, T-even
interaction
that turns out to be enhanced by $\sim A^{1/3}$ times (see \cite{IPNCI}).

The effects of renormalization of P-odd, T-even interaction
were considered in details in Ref. \cite{IPNCI}; below, we focus our
attention on the T-,P-odd interaction.

\section{Explicit form of the resulting two-particle T-,P-odd
interaction}
\label{sec:level4}

To solve the equation (8) and find an explicit form of the
ITPNCI we use
the Landau-Migdal interaction \cite{Landau},\cite{Migdal},\cite{Brown}.
It is the most widely used particle-hole
interaction of contact type with spin- and isospin-exchange
terms which goes backwards
to Landau Fermi liquid theory (Ref.\cite{Landau}); for the
case of a nucleus it
was established in the Theory of Finite Fermi Systems
\cite{Migdal,Brown,SAPERSHTEIN} by
summation of all
graphs irreducible in the particle-hole direction.
This interaction can be written explicitly as follows
\begin{equation}
V({\bf r}_{1},\vec{\sigma}_{1}{\bf r}_{2},\vec{\sigma}_{2})=
C\delta({\bf r}_{1}-{\bf r}_{2})[f+f'{\bf \tau}_{1}{\bf \tau}_{2}+
g{\bf \sigma}_{1}{\bf \sigma}_{2}+g'{\bf \tau}_{1}{\bf \tau}_{2}
{\bf \sigma}_{1}{\bf \sigma}_{2}],
\end{equation}
where $C=\frac{\pi^{2}}{p_{F}m}=300$ $MeV \times fm^{3}$ is the universal
Migdal constant \cite{Migdal,Brown,SAPERSHTEIN} and the
strengths $f,f',g,g'$ are in fact functions of $r$ via density dependence:
$f=f_{in}-(f_{ex}-f_{in})(\rho(r)-\rho(0))/\rho(0)$ (the same for $f',g,g'$).
(Quantities subscripted by ``in'' and ``ex''
characterize interaction strengths in the depth of the nucleus and on
its surface, respectively).
With its parameter values
listed below, this interaction has been successfully used by many authors
(see Refs. \cite{Brown}) to quantitatively describe many
properties of heavy nuclei.

The conventional choice of the constants widely used for heavy nuclei is
(see \cite{Migdal,Brown,SAPERSHTEIN}):
$f_{ex}=-1.95$, $f_{in}=-0.075$, $f'_{ex}=0.05$, $f'_{in}=0.675$,
$g_{in}=g_{ex}=0.575$, and $g'_{in}=g'_{ex}=0.725$.

It can be seen that, in the same approximation of constant density
as used above, the operator ${\hat {\cal A}}$ is proportional to ${\hat
\alpha}$:
${\hat {\cal A}} = i {\tilde \theta} (\vec{\sigma} \vec{\nabla})$.
Evaluating the commutator in (13,14), we obtain
\begin{eqnarray}
[{\hat {\cal A}},V_{S}] =
\tilde{\theta}_{1}C(f+f'{\bf \tau}_{1} {\bf \tau}_{2}) \quad \vec{\sigma}_{1}
[\vec{\nabla}_{1},\delta(\vec{r}_{1}-\vec{r}_{2})] +
\nonumber\\
+\tilde{\theta}_{1}C(g+g'{\bf \tau}_{1} {\bf \tau}_{2}) \quad \vec{\sigma}_{2}
[\vec{\nabla}_{1},\delta(\vec{r}_{1}-\vec{r}_{2})]
\nonumber\\
+\tilde{\theta}_{2}C(f+f'{\bf \tau}_{1} {\bf \tau}_{2}) \quad \vec{\sigma}_{2}
[\vec{\nabla}_{2},\delta(\vec{r}_{1}-\vec{r}_{2})]
\nonumber\\
+\tilde{\theta}_{2}C(g+g'{\bf \tau}_{1} {\bf \tau}_{2}) \quad \vec{\sigma}_{1}
[\vec{\nabla}_{2},\delta(\vec{r}_{1}-\vec{r}_{2})]
\nonumber\\
-i C(g+g'{\bf \tau}_{1} {\bf \tau}_{2})
\vec{\sigma}_{1} \times \vec{\sigma}_{2}
\{ {\tilde \theta}_{1} \vec{\nabla}_{1}-
{\tilde \theta}_{2} \vec{\nabla}_{2}, \delta(\vec{r}_{1}-\vec{r}_{2}) \}.
\end{eqnarray}
Contrary to the case of P-odd, T-even weak interaction \cite{PION},
\cite{IPNCI}, averaging over the core nucleons here yields
nonzero result
\begin{displaymath}
\langle [{\hat {\cal A}},V_{S}] \rangle \neq 0,
\end{displaymath}
and, consequently, gives a nonzero contribution to the ``compensation
equation'' (14).
Taking together the terms with the same operator structures, we
obtain from (14) and (17) equations of type \cite{PION}
\begin{equation}
{\tilde \theta} (\vec{\sigma} \vec{\nabla}) U =
\theta (\vec{\sigma} \vec{\nabla}) U +
\gamma \frac{\rho(0)}{U(0)}
(\vec{\sigma} \vec{\nabla}) U.
\end{equation}
which is equivalent to a system of two linear algebraic equation
relating new (renormalized) interaction strengths
$\tilde{\eta}_{1,2}$ with their initial values  $\eta_{12}$
(without strong interaction).
The solutions for this system of equations
for
the constants are the following:
\begin{eqnarray}
{\tilde \eta}_{p}=
\frac{1}{D} \Bigl[ \left(1+ \tilde{C} g_{pp}\frac{N}{A} \right)
\left(\frac{Z}{A} \eta_{pp} + \frac{N}{A} \eta_{pn} \right)
\nonumber\\
-
\tilde{C} g_{pn} \frac{N}{A} \left( \frac{N}{A} \eta_{np} +
\frac{Z}{A} \eta_{pp} \right) \Bigr],
\nonumber\\
{\tilde \eta}_{n}=
\frac{1}{D} \Bigl[ \left(1+ \tilde{C} g_{pp} \frac{Z}{A} \right)
\left( \frac{N}{A} \eta_{nn} + \frac{Z}{A} \eta_{np} \right)
\nonumber\\
-
\tilde{C} g_{pn} \frac{Z}{A} \left( \frac{Z}{A} \eta_{pp} +
\frac{N}{A} \eta_{pn} \right) \Bigr],
\end{eqnarray}
with
$D=1+ \tilde{C} g_{pp} +
4 \tilde{C}^{2} g_{pn}^{2}
Z N/A^{2}$.
Here, ${\tilde C} = C \rho / |U| = \frac{4}{3} \frac
{\varepsilon_{F}}{|U|}=
\frac{4}{3}(1+\frac{B_{n}}{\varepsilon_{F}})^{-1}
\simeq 1$ and
$\eta^{0}_{p}$ and $\eta^{0}_{n}$ are
the initial values of the constants. We have used the well known
relations :
\begin{equation}
C=\frac{\pi^{2}}{p_{F} m}, \quad \rho= \frac{2p_{F}^{3}}
{3\pi^{2}}, \quad
\varepsilon_{F}=\frac{p_{F}^{2}}{2m}, \quad |U|= \varepsilon_{F}+
B_{n},
\end{equation}
where $p_{F}$ is a Fermi momentum, $B_{n}$ is a nucleon
separation
energy.
The renormalized matrix elements of the T-,P-odd weak potential
for $^{209}Pb$ are presented in the last column of Tables 1,2.
It is seen that
the strong residual interaction reduces the values of the
T-,P-odd
potential constants $1.5...2$ times, as average.

To this end, from (15) with account for (17) we obtain
the resulting purely two-body T-,P-odd weak interaction
in a nucleus that can be written as following:
\begin{eqnarray}
{\cal W}^{{\small P,T}}_{eff} \quad = \quad
:{\cal W}^{{\small P,T}}: \quad + \quad {\cal W}^{{\small P,T}}_{IPTNCI}=
\quad \frac{G}{\sqrt{2}} \frac{1}{2m}
\qquad \nonumber\\
\times \Bigl[ :
\left( (\eta_{12}+ \tilde \eta_{2} g_{12} \tilde C) \vec{\sigma}_{1}-
(\eta_{21} - \tilde{\eta}_{1} g_{12} \tilde{C}) \vec{\sigma}_{2} \right)
[\vec{\nabla}_{1},\delta(\vec{r}_{1}-\vec{r}_{2})]
\qquad \nonumber\\
+ \tilde C (\tilde \eta_{1} \vec{\sigma}_{1} -
\tilde \eta_{2} \vec{\sigma}_{2})
[\vec{\nabla}_{1},\delta(\vec{r}_{1}-\vec{r}_{2})f_{12}(r_{1})]: \nonumber\\
-i \quad
\vec{\sigma}_{1} \times \vec{\sigma}_{2}
\{ ({\tilde \eta}_{1}g_{12} \tilde C +\eta'_{12})
\vec{\nabla}_{1}-
({\tilde \eta}_{2}g_{12} \tilde C +\eta'_{12})
\vec{\nabla}_{2}, \delta(\vec{r}_{1}-\vec{r}_{2}) \}
\Bigr],
\qquad \qquad \qquad
\end{eqnarray}
where the constants $\eta_{1},\eta_{2}$ are given by Eq. (19).
We used here the fact that spin constant of the strong interaction
(16) does not depend on $r$, while the constants $f_{pp}=f_{nn}=f(r)+f'(r)$,
$f_{pn}=f_{np}=f(r)-f'(r)$ do.

It should be noted that the induced T-,P-odd interaction
${\cal W}^{TP}_{ITPNCI}$ has the same operator structure
as the initial two-body T-,P-odd interaction ${\cal W}^{TP}$.
Thus, ${\cal W}^{TP}_{ITPNCI}$ differs from ${\cal W}^{TP}$
only due to renormalization of strength constants which
turns out to be weak, because
response of the
nucleus to the T- and P-odd potential (4) as a function of
the interaction constants
has poles ($D=0$) at $g={\tilde C}^{-1} \simeq -1$ and $g' \simeq
{\tilde C}^{-1} \simeq -1$ (for $N \simeq Z$), while
the actual nuclear strong interaction ``drives''
the solution of the renormalization equations (18) to the direction
opposite to poles.
As a result, induced T-,P-odd interaction does not play especial
role in the present case and causes renormalization of
order $1 \div 2$.
Thus there is essential difference with
the case of the P-odd, T-even
weak interaction \cite{IPNCI} where the analogous induced P-odd interaction
is enhanced by abou $A^{1/3}$ times and practically dominates
the results.

In practical calculations, it is convenient
to treat ${\cal W}^{TP}_{eff}$ in the
secondly quantized version
using multipole expansion in the particle-hole channel:
${\cal W}^{TP}_{eff}=
\frac{1}{2}\sum_{J}((a^{+}b)_{J}{\cal W}^{TP,J}_{eff \quad
abcd}(c^{+}d)_{J})_{0}$
where
$(...)_{J}$ means the coupling of nucleon creators $a^{\dagger}$ and
destructors $a$ to a given angular momentum $J$ \cite{BM}.
The numerical results for some reduced matrix elements of ${\cal
W}^{TP,J}_{eff}$
as compared to those of the initial interaction $:{\cal W}^{TP,J}:$
between valence
shell states for Th-U region are presented in the Table 3.

\section{T-,P-odd matrix elements between compound states}
\label{sec:level5}

In the work (Ref. \cite{we}) we have introduced a method to calculate
Mean Squared Matrix Elements (MSME) of operators between compound states
and have obtained the results for P-odd, T-even weak interaction.
Here, we apply this method to calculation of MSME of P-,T-odd
interaction.
Consider the mean squared value of this matrix element:
\begin{eqnarray}
\overline{ { {\cal W}^{P,T} }^{2}} \quad = \quad
\overline{(p|{\cal W}^{P,T}|s)(s|{\cal W}^{P,T}|p)} \quad =
\qquad \qquad \qquad
\nonumber\\
\quad \overline{(p|:{\cal W}^{P,T}:+
{\cal W}^{P,T}_{ITPNCI}|s)
(s|:{\cal W}^{P,T}:+  {\cal W}^{P,T}_{ITPNCI}   |p)}
\end{eqnarray}
We can expand now the compound states $|C^{J \pi}) =|s),|p)$ in terms
of their simple components (multiparticle excitations) $|\alpha^{J \pi}>$
of the same quantum numbers of angular momentum $J$ and parity $\pi$,
\begin{equation}
|C) \quad = \quad \sum_{\alpha} C_{\alpha} |\alpha>,
\end{equation}
having for the MSME the expression
\begin{equation}
\overline{ { {\cal W}^{P,T} }^{2}} \quad = \quad
\sum_{\alpha \beta} \overline{C_{\alpha}C_{\beta}
(p|:{\cal W}^{P,T}:+  {\cal W}^{P,T}_{ITPNCI}|\alpha>
<\beta|:{\cal W}^{P,T}:+  {\cal W}^{P,T}_{ITPNCI}|p)}
\end{equation}
The number of different terms in the Eq.(22), ${\cal N}$, is very
large $\sim 10^{5} \div 10^{7}$. The main contribution in Eq.(22)
is dominated
by the set of ${\bar N}$ ``principal components'' $|{\bar \alpha}>$
with shell-model energies $E_{{\bar \alpha}}$ close to the
energy of a compound state $E$.
We can make use of the statistical independence of the coefficients
$C_{\alpha}$ to take their second moments in the form
(Ref.\cite{BM},
\cite{SF}):
\begin{equation}
\overline{C_{\alpha}C_{\beta}}\quad =\quad \overline{C^{2}_{\alpha}}
\delta_{\alpha \beta}\quad=
\quad \delta_{\alpha \beta} \frac{1}{\overline{N}}
\Delta(\Gamma_{spr},E-E_{\alpha}).
\end{equation}
Bar means the averaging over a rather broad set of the
compound states.
Here, the spreading width $\Gamma_{spr}$ is related to the
number of principal components
$\overline{N}^{-1/2} \simeq \sqrt{\frac{2d}{\pi \Gamma_{spr}}}$
and $d$ is the average energy distance between the resonances.
The Breit-Wigner-type factor $\Delta$, describing cutting off of weights
before states distanced in energy,
\begin{equation}
\Delta(\Gamma_{spr},E-E_{\alpha})=
\frac{\Gamma_{spr}^{2}/4}{(E-E_{\alpha})^{2}+\Gamma_{spr}^{2}/4},
\end{equation}
may be treated as a ``spread''
$\delta$-function. It is normalized as to be of
order unity for $|E-E_{\alpha}|\leq\Gamma_{spr}/2$ and with conventional
limit $\Delta(\Gamma_{spr},E-E_{\alpha})
\to$ $\frac{\pi\Gamma_{spr}}{2}\delta(E-E_{\alpha})$ for $\Gamma_{spr}\to 0$.
For the principal components, $|E_{\alpha}-E| \alt \Gamma_{spr}$,
expression (25) reflects ``chaotic'' nature of a broad mixture of the simple
components in the compound state due to the strong interaction.
For the small (energy distanced) components it reduces to the
perturbation theory result.
{}From (23)-(25), we obtain for the MSME
\begin{eqnarray}
\overline{ { {\cal W}^{P,T} }^{2}}  \quad
= \quad \sum_{\alpha} \frac{1}{\overline{N}}
\Delta(\Gamma_{spr},E-E_{\alpha})
\nonumber\\
\overline{(p|:{\cal W}^{P,T}:+  {\cal W}^{P,T}_{ITPNCI}|\alpha>
<\alpha|:{\cal W}^{P,T}:+  {\cal W}^{P,T}_{ITPNCI}|p)}.
\end{eqnarray}
The argument of the function $\Delta$ here is the change of the
energy: $E-E_{\alpha}=\epsilon_{a}-\epsilon_{b}+\epsilon_{c}-\epsilon_{d}$,
and $\tilde{V}$ is given by Eq.(6).
Summation over $\alpha$ in (26)
is equivalent to summation over different components of the operator
${\cal W}^{P,T}_{eff}$
in Eq.(5), i.e. the problem is reduced
to the calculation of $(p|{\cal W}^{P,T} {\cal W}^{P,T}|p)$.
The
coefficients before the ``principal'' components $\tilde C_{\alpha}$
in (22) are governed by the microcanonical ensemble rule \cite{BM,SF}.
Then, to calculate the averaging over p-resonance ``principal'' components
$\overline{(p| ... |p)}$ in ${\overline {\cal W}^{P,T 2}}$,
we use, instead
of the present microcanonical ensemble, an equivalent canonical one.
The latter can always be introduced
for a system with a large number degrees of freedom by
introducing the effective nuclear temperature T and chemical potentials
$\lambda_{n},\lambda_{p}$.
In the second quantization representation,
the average expectation value
in (26) is reduced to a
canonical ensemble average with the standard contractor rules
$(p|\overline{a^{+}b}|p)=\delta_{ab}\nu^{T}_{a}$, for
$\nu^{T}_{a}$ being the finite temperature Fermi occupation probabilities,
$\nu^{T}_{a}=\lbrace exp[(\epsilon_{a}-\lambda)/T]+1
\rbrace ^{-1}$. The canonical ensemble parameters $T$, $\lambda_{\tau}$ ($\tau$
means isospin projection) are to be determined from conventional ``consistency
`` equations $E = \sum_{a}\nu_{a}\epsilon_
{a}$, $ Z=\sum_{p}\nu_{p}$, and $N=\sum_{n}\nu_{n}$
for the excitation energy $E$ (being equal the to neutron separation energy,
$B_{N}$),
nuclear charge $Z$, and neutron number $N$ correspondingly.

By means of the same considerations, we obtain
the following result for $\sqrt{\overline{ { {\cal W}^{{\small P,T}} }^{2}}}$:
\begin{eqnarray}
\sqrt{\overline{ { {\cal W}^{{\small P,T}} }^{2}}}
= \sqrt{\frac{2d}{\pi\Gamma_{spr}}}
\Bigl\{ \frac{1}{2} \sum_{abcd}\nu^{T}_{a}(1-\nu^{T}_{b})
\nu^{T}_{c}(1-\nu^{T}_{d})
\nonumber\\
\mid {\cal W}^{{\small P,T}}_{eff \quad ab,cd}
\mid ^{2}\Delta(\Gamma_{spr},
\epsilon_{a}-\epsilon_{b}+\epsilon_{c}-\epsilon_{d})\Bigl\} ^{\frac{1}{2}}.
\end{eqnarray}
Here, $\Delta(\Gamma_{spr},
\varepsilon_{a}-\varepsilon_{b}+\varepsilon_{c}-\varepsilon_{d})$
can be viewed as an approximate energy
conservation law with the accuracy up to width of states.

The numerical calculations for $^{233}$Th have been performed with the use
of single-particle basis of states obtained by numerically
(see Sec.II, Eq.(7) and below).

The value of temperature $T=0.6MeV$ was used in accordance to the consistency
condition for excitation energy.
The result for the mean squared matrix elements of T-,P-odd interaction
between compound states is
\begin{displaymath}
\sqrt{\overline{ { {\cal W}^{{\small P,T}} }^{2}}}=
0.20 \eta_{0} meV.
\end{displaymath}
The ratio of the P,T-odd matrix elements to the P-odd ones
is $\sqrt{\overline{ { {\cal W}^{P,T} }^{2}}} /
\sqrt{\overline{ { {\cal W}^{P} }^{2}}} = 0.1 \eta / g$.
Here, we use equal values of constants $\eta_{12}$ in (2),
$\eta_{12}=\eta_{0}$.
The corresponding mixing coefficient for compound states $|F_{sp}|$
is
\begin{displaymath}
|F_{sp}| \simeq \frac{\sqrt{\overline{ { {\cal W}^{P,T} }^{2}}}}
{|E_{s}-E_{p}|}  \simeq 1. \cdot 10^{-5} \eta_{0}
\end{displaymath}
that is about $10^{3}$ times larger than single-particle mixing $f_{12} \simeq
10^{-8} \eta_{0}$ (Eq.(8)).
We assumed in this estimate that $|E_{s}-E_{p}|=D_{s}$
where $D_{s}$ is the average energy interval between compound
resonances in $s$-wave.

\section{Correlations between T-,P-odd and P-odd, T-even matrix
in compound states}
\label{sec:level6}

The question on possible correlations between matrix elements
of P-,T-odd weak interaction and those of P-odd, T-even
weak interaction is very interesting.
Knowing the correlator
\begin{equation}
C(W^{P},{\cal W}^{P,T})
= \frac{
\overline{ (p| W^{P} |s) (p|{\cal W}^{P,T}|s) }
}
{
\sqrt{\overline{ { {\cal W}^{P,T} }^{2}}}
\sqrt{\overline{ { W^{P} }^{2}}}
}
\end{equation}
($0 < |C(W^{P},{\cal W}^{P,T})| < 1$),
one can make predictive estimates on the values and signs of the
P-,T-odd effects in compound states basing on the information
about the corresponding quantities for P-odd effects
(the latter are much easier to be measured)
in the case when the quantity
$C(W^{P},{\cal W}^{P,T})$
differs considerably from zero.
$C(W^{P},{\cal W}^{P,T})$
can be calculated, in principle, by the same technique \cite{we}
as mean squared matrix element \cite{FLAMBAUM}. We can employ
for the calculation of
the numerator of Eq.(29)
$c(W^{P},{\cal W}^{P,T})=
\overline{ (p| W^{P} |s) (p|{\cal W}^{P,T}|s) }$
the method of reducing to averaging over the
ensemble
\begin{eqnarray}
c(W^{P},{\cal W}^{P,T})=
\overline{ (p|W^{P}|s)(s|{\cal W}^{P,T}|p)}  \quad =
\nonumber\\
\quad \overline{(p|
W^{P}_{IPNCI}|s)
(s|:{\cal W}^{P,T}:+
{\cal W}^{P,T}_{ITPNCI}
|p)}
\end{eqnarray}
However, in this case more careful treatment is needed.
The present thermodynamical approach makes no difference between
the cases when ``external'' averaging (canonical) goes
over p-states either s-states provided the mean squared
matrix elements are not very sufficient to change
$\overline{ (p|W^{P}|s)(s|W^{P}|p) }
\rightarrow
\overline{ (s|W^{P}|p)(p|W^{P}|s) }$.
The latter is not the case for the quantity
$c(W^{P},{\cal W}^{P,T})$. The reason is that
the matrix elements of P-odd weak operator
$W^{P}$ are imaginary and change sign when
substituting final states instead of initial states.
On the contrary, the matrix elements of P-,T-odd weak operator
${\cal W}^{P,T}$ are real and symmetric under
such substitution.
Two-body matrix elements of $W^{P}$ and ${\cal W}^{P,T}$
obey the following symmetry rules respectively:
\begin{eqnarray}
W^{P}_{ab,cd} \quad = -W^{P}_{ba,dc} \quad =
-W^{P}_{dc,ba} \quad = W^{P}_{cd,ab}  \nonumber\\
{\cal W}^{P,T}_{ab,cd} \quad =
{\cal W}^{P,T}_{ba,dc} \quad =
{\cal W}^{P,T}_{dc,ba} \quad =
{\cal W}^{P,T}_{cd,ab}
\end{eqnarray}
As a result, we have some cancellations of the different terms
in the sum of the products ${\cal W}^{P,T} W^{P}$.
Thus, a symmetrization should be done when
the quantity $\bar{C}(W^{P},{\cal W}^{P,T})$
is calculated by the present method of reduction to the ensemble averaging:
\begin{eqnarray}
c(W^{P},{\cal W}^{P,T})=
\nonumber\\
\frac{1}{2} \left[ \overline{(p|
{\cal W}^{{\small P,T}}_{eff}|s)
(s|W^{{\small P}}_{IPNCI}|p)} +
\overline{(s|
W^{{\small P}}_{IPNCI}|p)
(p|{\cal W}^{{\small P,T}}_{eff}|s)} \right] =
\nonumber\\
=\frac{1}{2} \left[ \overline{(p|
W^{{\small P}}_{IPNCI}|s)
(s|{\cal W}^{{\small P,T}}_{eff}|p)} -
\overline{(s|
W^{{\small P}}_{IPNCI}|p)
(p|{\cal W}^{{\small P,T}}_{eff}|s)} \right]
\end{eqnarray}
As a result of symmetrization and the negative sign before
the second term in the last line, some cancellations
of similar terms in the large sum of the same type as in
Eq.(23) are possible.
{}From the last equation, it is seen that we can not
pretend to obtain the correct sign of the correlator
within present statistical method, because
the compound states of positive and negative parity are
treated on the same footing. Without having an
additional information about occupancies of particular
single-particle levels with a given total angular momentum
and parity, only absolute valu of the correlator can
be estimated.

After evaluation of commutator in Eq.(30) and thermal averaging
we obtain the following expression for the numerator in (32)
\begin{eqnarray}
|c(W^{{\small P}},{\cal W}^{{\small P,T}})|=
\frac{d}{\pi\Gamma_{spr}}
\mid \sum_{abcd}  \nu^{T}_{a}(1-\nu^{T}_{b})\nu^{T}_{c}(1-\nu^{T}_{d})
W^{{\small P}}_{ab,cd}
{\cal W}^{{\small P,T}}_{eff \quad dc,ba} \mid
\nonumber\\
\times \Delta(\Gamma_{spr},
\epsilon_{a}-\epsilon_{b}+\epsilon_{c}-\epsilon_{d}).
\end{eqnarray}
Using this result and equations (24),(25,(26) we obtain, for the same
value of temperature and the same single-particle basis as in
the calculations of mean squared matrix elements, the following absolute
value of
the correlator (26) for the $^{233}Th$:
\begin{displaymath}
\mid C(W^{{\small P}},{\cal W}^{{\small P,T}}) \mid \quad \simeq
\quad 0.1 .
\end{displaymath}
It means that correlations in matrix elements are weak. Of course, the
present statistical calculation is estimative, and to obtain more
definite result for correlator more refined technique is needed.

\section{Summary}
\label{sec:level7}

To conclude, we have considered the T-,P-odd nucleon
interaction in heavy nuclei.
Effects of the renormalization of
this interaction are considered.
An effective two-body T,P-odd
interaction acting near the Fermi surface
is obtained and its matrix elements are calculated.
This interaction accumulates the effects of the distant
states admixtures.
We obtained the results for means squared values of T- and P-violating
interaction between compound states of opposite parity. As well
as in the case of P-odd, T-even weak interaction, statistical
enhancement of T-,P-odd effects in neutron resonances take
place. The enhancement is about $10^{3}$ times for the mixing between
compound states of opposite parity as compared to
the single-particle mixing.

Correlations between matrix elements of T-,P-odd and P-odd, T-even
interactions in compound states are found to be weak within the
statistical model.

\section{Aknowledgement}
We are grateful to V.F.Dmitriev and V.B.Telitsin
kindly providing us with the code for the nuclear wave functions
calculation.


\begin{table}
\caption{Single-particle matrix elements of P,T-odd weak interaction
for protons.
The levels closest to the Fermi energy are
marked by asterisks.
}

\begin{tabular}{cccccc}
a & b
& $e_{a}-e_{b}$ & $w^{P}_{ab}$ & $w^{TP}_{ab}$
& ${\tilde w^{TP}}_{ab}$ \\
 &  & (MeV) & eV & eV & eV \\
\tableline
$ 2 p_{  3/2}$ & $  2  d_{  3/2}^{*}$ &  -8.554 &    0.513$g_{pp}+$
0.748$g_{pn}$ &
$ 0.080\eta_{pp}+ 0.098\eta_{pn}$ &
$0.053 \eta_{pp}+ 0.065\eta_{pn}$\\
& & & & &  $+0.003\eta_{np}+   0.003\eta_{nn}$\\
$ 1 g_{  9/2}$ & $  1  h_{  9/2}^{*}$ & -11.054 &
0.599$g_{pp}+$   0.842$g_{pn}$ &  $ 0.112\eta_{pp}+
0.129\eta_{pn}$ &
$0.074\eta_{pp}+   0.086\eta_{pn}$\\
$ 2 p_{  1/2}$ & $  3  s_{  1/2}^{*}$ &  -8.443 &
-0.500$g_{pp}$  -0.722$g_{pn}$ &  -0.066$\eta_{pp}$
-0.078$\eta_{pn}$ &
$-0.044\eta_{pp}  -0.052\eta_{pn}$\\
$ 2 p_{  1/2}$ & $  3  s_{  1/2}^{*}$ &  -8.443 &
-0.500$g_{pp}$  -0.722$g_{pn}$ &  -0.066$\eta_{pp}$
-0.078$\eta_{pn}$ &
$-0.044\eta_{pp}  -0.052\eta_{pn}$\\
$ 1 g_{  7/2}$ & $  2  f_{  7/2}$ &  -9.745 &
-0.517$g_{pp}$  -0.720$g_{pn}$ &  -0.068$\eta_{pp}$
-0.071$\eta_{pn}$ &
$-0.045\eta_{pp}  -0.047\eta_{pn}$\\
$ 2 d_{  5/2}$ & $  2  f_{  5/2}$ & -10.084 &
0.553$g_{pp}+$   0.812$g_{pn}$ &   0.078$\eta_{pp}+$
   0.107$\eta_{pn}$ &
$ 0.052\eta_{pp}+   0.071\eta_{pn}$\\
$ 2 d_{  3/2}^{*}$ & $  2  p_{  3/2}$ &   8.554 &
-0.513$g_{pp}+$  -0.748$g_{pn}$ &   0.080$\eta_{pp}+$
   0.098$\eta_{pn}$ &
$0.053\eta_{pp}+   0.065\eta_{pn}$\\
& & & & & $  + 0.003\eta_{np}+   0.003\eta_{nn}$\\
$ 2 d_{  3/2}^{*}$ & $  3  p_{  3/2}$ &  -8.732 &
-0.558$g_{pp}$  -0.803$g_{pn}$ &  -0.050$\eta_{pp}$
  -0.073$\eta_{pn}$ &
$-0.033\eta_{pp}  -0.048\eta_{pn}$\\
$ 3 s_{  1/2}^{*}$ & $  3  p_{  1/2}$ &  -9.186 &
    0.549$g_{pp}+$   0.806$g_{pn}$ &
 0.055$\eta_{pp}+$   0.091$\eta_{pn}$ &
$0.037\eta_{pp}+ 0.060\eta_{pn}$\\
& & & & & $ +  0.002\eta_{np}+   0.003\eta_{nn}$\\
$ 1 h_{  9/2}^{*}$ & $  1  g_{  9/2}$ &  11.054 &
   -0.599$g_{pp}$  -0.842$g_{pn}$ &
0.112$\eta_{pp}+$   0.129$\eta_{pn}$ &
$0.074\eta_{pp}+   0.086\eta_{pn}$\\
$ 1 h_{  9/2}^{*}$ & $  2  g_{  9/2}$ &  -9.417 &
   -0.575$g_{pp}$  -0.789$g_{pn}$ &  -0.055$
\eta_{pp}$  -0.064$\eta_{pn}$ &  $-0.037\eta_{pp}
  -0.042\eta_{pn}$\\
$ 2 f_{7/2}$ & $  1  g_{  7/2}$ &   9.745 &
    0.517$g_{pp}+$   0.720$g_{pn}$ &  -0.068$
\eta_{pp}$  -0.071$\eta_{pn}$ & $-0.045\eta_{pp}
  -0.047\eta_{pn}$\\
$ 2 f_{5/2}$ & $  1  d_{  5/2}$ &  26.505 &
    0.096$g_{pp}+$   0.134$g_{pn}$ &  -0.067$
\eta_{pp}$  -0.066$\eta_{pn}$ & $-0.044\eta_{pp}
 -0.044\eta_{pn}$\\
$ 2 f_{  5/2}$ & $  2  d_{  5/2}$ &  10.084 &
   -0.553$g_{pp}$  -0.812$g_{pn}$ &   0.078$
\eta_{pp}+$   0.107$\eta_{pn}$ & $0.052\eta_{pp}+
   0.071\eta_{pn}$\\
$ 3 p_{  3/2}$ & $  1  d_{  3/2}$ &  25.840 &
   -0.054$g_{pp}$  -0.063$g_{pn}$ &   0.036$
\eta_{pp}+$   0.036$\eta_{pn}$ & $0.024\eta_{pp}+
   0.024\eta_{pn}$\\
$ 3 p_{  3/2}$ & $  2  d_{  3/2}^{*}$ &   8.732 &
    0.558$g_{pp}+$   0.803$g_{pn}$ &  -0.050$
\eta_{pp}$  -0.073$\eta_{pn}$ & $-0.033\eta_{pp}
  -0.048\eta_{pn}$\\
\end{tabular}
\end{table}
%
\narrowtext


\widetext
\begin{table}
\caption{The same as in Table 1, but for neutrons.}
\begin{tabular}{cccccc}
a & b
& $e_{a}-e_{b}$ & $w^{P}_{ab}$ & $w^{TP}_{ab}$
& ${\tilde w^{TP}}_{ab}$ \\
 &  & (MeV) & eV & eV & eV \\
\tableline
$ 3 p_{  3/2}^{*}$ & $  2  d_{  3/2}$ &   7.784 &    0.541$g_{nn}+$
0.778$g_{np}$ &  -0.071$\eta_{nn}$
  -0.048$\eta_{np}$ &  $-0.040\eta_{nn}
-0.027\eta_{np}$\\
$ 3 p_{  3/2}^{*}$ & $  3  d_{  3/2}$ &  -8.733 &    0.446$g_{nn}+$
0.661$g_{np}$ &   0.060$\eta_{nn}+$   0.026$\eta_{np}$ & $0.033\eta_{nn}+
0.015\eta_{np}$\\
$ 2 f_{  5/2}^{*}$ & $  2  d_{  5/2}$ &  10.055 &
   -0.536$g_{nn}$  -0.790$g_{np}$ &   0.108$\eta_{nn}+$   0.078$\eta_{np}$ &
$0.060\eta_{nn}+   0.044\eta_{np}$\\
$ 2 f_{  5/2}^{*}$ & $  3  d_{  5/2}$ &  -6.633 &
   -0.539$g_{nn}$  -0.773$g_{np}$ &  -0.044$
\eta_{nn}$  -0.017$\eta_{np}$ & $-0.024
\eta_{nn}  -0.009\eta_{np}$\\
$ 2 f_{  5/2}^{*}$ & $  2  d_{  5/2}$ &  10.055 &   -0.536$g_{nn}$
-0.790$g_{np}$ &
0.108$\eta_{nn}+$   0.078$\eta_{np}$ & $0.060\eta_{nn}+   0.044\eta_{np}$\\
$ 2 f_{  5/2}^{*}$ & $  3  d_{  5/2}$ &  -6.633 &
   -0.539$g_{nn}$  -0.773$g_{np}$ &
  -0.044$\eta_{nn}$  -0.017$\eta_{np}$ &$-0.024\eta_{nn}  -0.009\eta_{np}$\\
$ 3 p_{  1/2}^{*}$ & $  1  s_{  1/2}$ &  34.531 &
    0.004$g_{nn}$  -0.003$g_{np}$ &   0.023$
\eta_{nn}+$   0.022$\eta_{np}$ &
$0.013\eta_{nn}+   0.012\eta_{np}$\\
& & & & &  $ 0.001\eta_{pn}+   0.001\eta_{pp}$\\
$ 3 p_{  1/2}^{*}$ & $  2  s_{  1/2}$ &  23.964 &
    0.037$g_{nn}+$   0.050$g_{np}$
&  -0.044$\eta_{nn}$  -0.042$\eta_{np}$ &
$-0.025\eta_{nn}  -0.023\eta_{np}$\\
& & & & & $-0.001\eta_{pn}  -0.001\eta_{pp}$\\
$ 3 p_{  1/2}^{*}$ & $  3  s_{  1/2}$ &   8.811 &
   -0.528$g_{nn}$  -0.775$g_{np}$ &
0.090$\eta_{nn}+$   0.054$\eta_{np}$ &
$0.050\eta_{nn}+   0.030\eta_{np}$\\
&  & & & & $0.002\eta_{pn}+   0.001\eta_{pp}$\\
$ 3 p_{  1/2}^{*}$ & $  4  s_{  1/2}$ &  -6.645 &
   -0.452$g_{nn}$  -0.660$g_{np}$ &  -0.035$
\eta_{nn}$  -0.012$\eta_{np}$ &  $-0.020\eta_{nn}
  -0.007\eta_{np}$\\
$ 2 g_{  9/2}$ & $  1  h_{  9/2}$ &
   7.590 &    0.561$g_{nn}+$   0.770$g_{np}$ &
  -0.058$\eta_{nn}$  -0.048$\eta_{np}$ &
$-0.033\eta_{nn}  -0.027\eta_{np}$\\
$ 3 d_{  5/2}$ & $  1  f_{  5/2}$ &  24.931 &    0.023$g_{nn}+$   0.039$g_{np}$
&   0.012$\eta_{nn}+$   0.004$\eta_{np}$ & $0.006\eta_{nn}+   0.002\eta_{np}$\\
$ 3 d_{  5/2}$ & $  2  f_{  5/2}^{*}$ &
   6.633 &    0.539$g_{nn}+$   0.773$g_{np}$
&  -0.044$\eta_{nn}$  -0.017$\eta_{np}$ &
$-0.024\eta_{nn}  -0.003\eta_{np}$\\
$ 4 s_{  1/2}$ & $  1  p_{  1/2}$ &  37.016 &
   -0.038$g_{nn}$  -0.040$g_{np}$ &
0.001$\eta_{nn}+$   0.006$\eta_{np}$ &
 $0.000\eta_{nn}+   0.003\eta_{np}$\\
$ 4 s_{  1/2}$ & $  2  p_{  1/2}$ &  23.364 &
    0.039$g_{nn}+$   0.053$g_{np}$ &   0.009$
\eta_{nn}$  -0.004$\eta_{np}$ & $0.005
\eta_{nn}  -0.002\eta_{np}$\\
$ 4 s_{  1/2}$ & $  3  p_{  1/2}^{*}$ &
   6.645 &    0.452$g_{nn}+$   0.660$g_{np}$
 &  -0.035$\eta_{nn}$  -0.012$\eta_{np}$
& $-0.020\eta_{nn}  -0.007\eta_{np}$\\
$ 2 g_{  7/2}$ & $  1  f_{  7/2}$ &  28.563
&    0.034$g_{nn}+$   0.068$g_{np}$ &
-0.066$\eta_{nn}$  -0.061$\eta_{np}$ &
$-0.037\eta_{nn}  -0.034\eta_{np}$\\
$ 2 g_{  7/2}$ & $  2  f_{  7/2}$ &  11.326
&   -0.518$g_{nn}$  -0.770$g_{np}$ &
0.101$\eta_{nn}+$   0.061$\eta_{np}$ &
$ 0.057\eta_{nn}+   0.034\eta_{np}$\\
$ 3 d_{  3/2}$ & $  2  p_{  3/2}$ &
25.118 &   -0.021$g_{nn}$  -0.020$g_{np}$
&  -0.028$\eta_{nn}$  -0.015$\eta_{np}$ &
$-0.016\eta_{nn}  -0.008\eta_{np}$\\
& & & & & $-0.001\eta_{pn}  -0.001\eta_{pp}$\\
$ 3 d_{  3/2}$ & $  3  p_{  3/2}^{*}$ &
8.733 &   -0.446$g_{nn}$  -0.661$g_{np}$
&   0.060$
\eta_{nn}+$   0.026$\eta_{np}$
& $0.033\eta_{nn}+   0.015\eta_{np}$\\
\end{tabular}
\end{table}
%

\narrowtext

\newpage

\widetext
\begin{table}
\caption{Reduced matrix elements of T-,P-odd
weak interaction ${\cal W}^{TP,J}_{abcd}$
for states of valence shells in the U-Th region
in terns of weak constants $\eta_{12}$ (Eq.(...)).
$a \equiv \{ p(n)n_{a}l_{a}j_{a} \}$, $p(n)$ means
proton (neutron) states. }
\begin{tabular}{cccccc}
$J$ & $a$ & $b$ & $c$ & $d$ &
${\cal W}^{TP,J}_{abcd}$ (eV) \\
\tableline
 2 & p$1h_{9/2}$ & p$1h_{9/2}$ & n$1j_{15/2}$ &
  bn$1i_{11/2}$ &  $ 0.021 \eta'_{pn}$ \\
 3 &  p$1h_{9/2}$ &   p$1h_{9/2}$ & n$1j_{15/2}$ &
n$1i_{11/2}$ &   -0.003$\eta_{pn}+  0.014\eta'_{pn}$ \\
 4 & p$1h_{9/2}$ &   p$1h_{9/2}$  &  n$1j_{15/2}$ &
n$1i_{11/2}$ &
  $ 0.010 \eta'_{pn}$ \\
 5 & p$1h_{9/2}$ &   p$1h_{9/2}$ & n$1j_{15/2}$ &
n$1i_{11/2}$ &
-0.006$\eta_{pn}+ 0.009\eta'_{pn}$\\
 6 &  p$1h_{9/2}$ & p$1h_{9/2}$ & n$1j_{15/2}$ &
 n$1i_{11/2}$ &
0.006$\eta'_{pn}$ \\
 7 &  p$1h_{9/2}$ &   p$1h_{9/2}$ & n$1j_{15/2}$ &
 n$1i_{11/2}$  &
$-0.011\eta_{pn}+   0.006\eta'_{pn}$ \\
 8 & p$1h_{9/2}$ & p$1h_{9/2}$ & n$1j_{15/2}$ &
 n$1i_{11/2}$ &
  0.004 $\eta'_{pn}$\\
 9 & p$1h_{9/2}$ & p$1h_{9/2}$ & n$1j_{15/2}$
& n$1i_{11/2}$  &
 $-0.024\eta_{pn}+    0.005\eta'_{pn}$ \\
 2&  n$1i_{11/2}$ &    n$1j_{15/2}$ &    n$2g_{9/2}$ &    n$2g_{9/2}$ &
$ -0.003\eta_{nn}  -0.009\eta'_{nn}$  \\
 3 & n$1i_{11/2}$ &    n$1j_{15/2}$ &    n$2g_{9/2}$ &    n$2g_{9/2}$ &
$  0.001\eta_{nn}   -0.006\eta'_{nn}$  \\
 4 & n$1i_{11/2}$ &    n$1j_{15/2}$ &    n$2g_{9/2}$ &    n$2g_{9/2}$ &
$ -0.002\eta_{nn}$    $-0.004\eta'_{nn}$  \\
 5 & n$1i_{11/2}$ &    n$1j_{15/2}$ &    n$2g_{9/2}$ &    n$2g_{9/2}$ &
$  0.002\eta_{nn}   -0.004\eta'_{nn}$ \\
 6 & n$1i_{11/2}$ &    n$1j_{15/2}$ &    n$2g_{9/2}$ &    n$2g_{9/2}$ &
 $-0.001\eta_{nn}   -0.003\eta'_{nn}$ \\
 7 & n$1i_{11/2}$ &    n$1j_{15/2}$ &    n$2g_{9/2}$ &    n$2g_{9/2}$ &
  $0.004\eta_{nn}   -0.003\eta'_{nn}$  \\
 8 & n$1i_{11/2}$ &    n$1j_{15/2}$ &    n$2g_{9/2}$ &    n$2g_{9/2}$ &
 $-0.001\eta_{nn}   -0.002\eta'_{nn}$ \\
 9 & n$1i_{11/2}$ &    n$1j_{15/2}$ &    n$2g_{9/2}$ &    n$2g_{9/2}$ &
  $0.009\eta_{nn}   -0.002\eta'_{nn}$ \\
 2 & n$1j_{15/2}$ &    n$1i_{11/2}$ &    n$1i_{11/2}$ &    n$2g_{9/2}$ &
 $-0.003\eta_{nn} -0.010\eta'_{nn}$  \\
 3 & n$1j_{15/2}$ &    n$1i_{11/2}$ &    n$1i_{11/2}$ &    n$2g_{9/2}$ &
  $0.002\eta_{nn}+  0.002\eta'_{nn}$\\
 4 & n$1j_{15/2}$ &    n$1i_{11/2}$ &    n$1i_{11/2}$ &    n$2g_{9/2}$ &
 $-0.003\eta_{nn}   -0.003\eta'_{nn}$\\
 5 & n$1j_{15/2}$ &    n$1i_{11/2}$ &    n$1i_{11/2}$ &    n$2g_{9/2}$ &
  $0.003\eta_{nn}+    0.001\eta'_{nn}$\\
 6 & n$1j_{15/2}$ &    n$1i_{11/2}$ &    n$1i_{11/2}$ &    n$2g_{9/2}$ &
 $-0.004\eta_{nn}   -0.001\eta'_{nn}$  \\
 7 & n$1j_{15/2}$ &    n$1i_{11/2}$ &    n$1i_{11/2}$ &    n$2g_{9/2}$ &
 $ 0.003\eta'_{nn}$  \\
 8 & n$1j_{15/2}$ &    n$1i_{11/2}$ &    n$1i_{11/2}$ &    n$2g_{9/2}$ &
 $-0.005\eta_{nn}   -0.001\eta'_{nn}$ \\
 9 & n$1j_{15/2}$ &    n$1i_{11/2}$ &    n$1i_{11/2}$ &    n$2g_{9/2}$ &
  $0.003\eta'_{nn}$ \\
\end{tabular}
\end{table}

\noindent
\end{document}